\documentclass[a4paper]{jpconf}

\usepackage{iwsmihead}
\usepackage{graphicx}
\usepackage{amssymb}

\begin{document}

\title{Spin glasses and algorithm benchmarks:\\A one-dimensional view}

\author{Helmut G.~Katzgraber\footnote[99]{Work done in collaboration with
W.~Barthel, S.~B\"ottcher, B.~Gon{\c c}alves, A.~K.~Hartmann,
M.~J\"unger, M.~K\"orner, F.~Krz\c{a}ka{\l}a, F.~Liers, D.~Sherrington
and A.~P.~Young.}}
\address{Theoretische Physik, ETH Z\"urich, CH-8093 Z\"urich, Switzerland}

\ead{katzgraber@phys.ethz.ch}

\begin{abstract}
Spin glasses are paradigmatic models that deliver concepts relevant
for a variety of systems. However, rigorous analytical results are
difficult to obtain for spin-glass models, in particular for realistic
short-range models. Therefore large-scale numerical simulations are
the tool of choice.  Concepts and algorithms derived from the study of
spin glasses have been applied to diverse fields in computer science
and physics.  In this work a one-dimensional long-range spin-glass
model with power-law interactions is discussed. The model has the
advantage over conventional systems in that by tuning the power-law
exponent of the interactions the effective space dimension can be
changed thus effectively allowing the study of large high-dimensional
spin-glass systems to address questions as diverse as the existence
of an Almeida-Thouless line, ultrametricity and chaos in short range
spin glasses.  Furthermore, because the range of interactions can be 
changed, the model is a formidable test-bed for optimization algorithms.
\end{abstract}

\section{Introduction}
\label{sec:introduction}

Spin glasses pose formidable challenges not only theoretically,
but also numerically \cite{binder:86}. Because analytically only the
mean-field Sherrington-Kirkpatrick (SK) model \cite{sherrington:75}
can be solved exactly, most of the research on realistic
short-range systems---such as the Edwards-Anderson Ising spin glass
\cite{edwards:75}---is performed numerically. Due to diverging
equilibration times in Monte Carlo simulations of spin glasses,
as well as an extra overhead because of configurational averaging, only
small systems can be studied. In order to probe the thermodynamic
limit it is therefore of paramount importance to use fast algorithms,
improved models, and large computer clusters.

Technological advances in the last decade have enabled the construction
of powerful multiprocessor machines out of commodity components at
low cost.  Still, the numerical effort required to study conventional
short-range spin glasses for low enough temperatures and large enough
system sizes exceeds the CPU time delivered by an average computer
cluster. Therefore, in addition to hardware advances, novel algorithms
need to be developed and tested, and improved models have to be used.

In this work we emphasize the importance of the choice of model when
studying spin glasses: the one-dimensional spin glass with power-law
interactions allows the study of large systems for effectively
high space dimensions. Furthermore, the model is an excellent
algorithm benchmark to test and improve modern algorithms to study
complex systems.  The model has the advantage, in that by tuning
the power-law exponent of the interactions the universality class
(effective space dimension) as well as the complexity of the system
can be changed. In what follows the model is introduced in detail. In
addition, past applications to the nature of the spin-glass state
\cite{katzgraber:03,katzgraber:03f}, ground-state energy distributions
in spin glasses \cite{katzgraber:04c} and the existence of a spin-glass
state in a field \cite{katzgraber:05c} are presented. Furthermore, new
applications to field chaos and ultrametricity in spin glasses, as well
as local-field distributions in spin glasses are presented. Finally,
future applications of the model to answer problems in the physics
of spin glasses are described, as well as applications to algorithm
development and testing.

\section{Model}
\label{sec:model}

\begin{figure}[h]
\includegraphics[width=18pc]{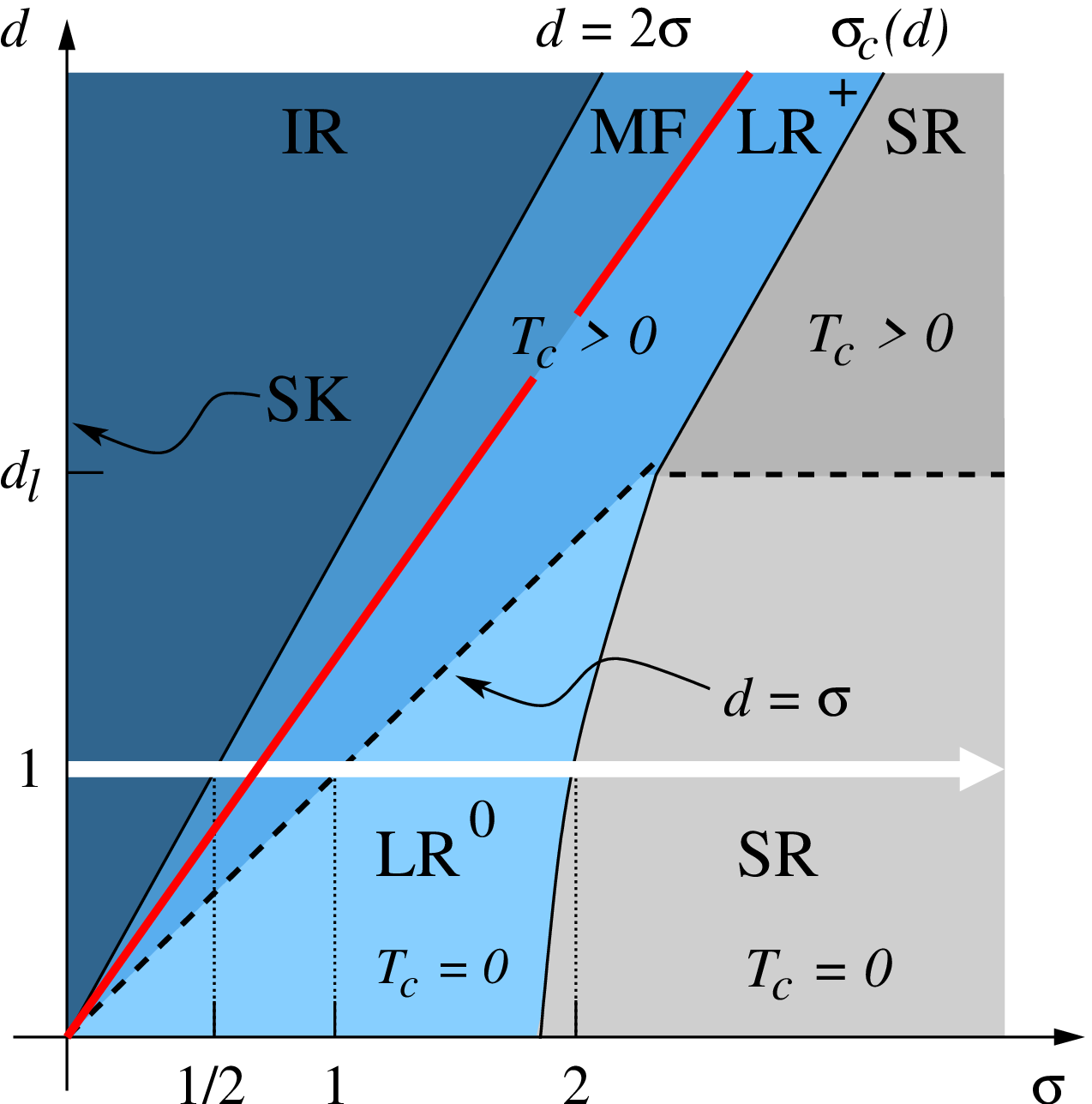}\hspace{2pc}
\begin{minipage}[b]{18pc}\caption{\label{fig:dsigma}

Sketch of the phase diagram in the $d$-$\sigma$ plane of the long-range
spin glass with power-law interactions. This work focuses only on
$d = 1$, which corresponds to the horizontal white arrow. By tuning
the power-law exponent $\sigma$ different universality classes
can be probed: For $\sigma \le 1/2$ ($d = 1$) the system is in
the infinite-range SK universality class. For $1/2 < \sigma \le
2/3$ the model exhibits a mean-field behavior corresponding to an
effective space dimension $d_{\rm eff} \ge 6$, where $d_{\rm eff}
\approx 2/(2\sigma - 1)$ for $1/2 \le \sigma \le 1$. The thick (red)
line separates mean-field from non-mean-field behavior.  For $2/3 <
\sigma < 1$ the model is a long-range spin glass with a finite ordering
temperature $T_{\rm c}$, whereas for $1 \le \sigma < 2$ the long-range
spin glass has $T_{\rm c} = 0$.  When $\sigma \ge 2$ [$\sigma_c(d)$]
the model is short-ranged with zero transition temperature. Figure
adapted from reference \cite{katzgraber:03}.

}
\end{minipage}
\end{figure}

The one-dimensional Ising spin glass with power-law interactions is given  by
the Hamiltonian
\cite{kotliar:83,bray:86b,fisher:88,katzgraber:03,katzgraber:03f,katzgraber:04c,katzgraber:05c}
\begin{equation}
{\mathcal H}= - \sum_{i<j} J_{ij} S_i S_j \, ,
\;\;\;\;\;\;\;\;\;\;
J_{ij}= c({\sigma}) \frac{\epsilon_{ij}}{{{r_{ij}}^\sigma}} \, ,
\;\;\;\;\;\;\;\;\;\;
r_{ij} = \frac{L}{\pi}\sin\left(\frac{\pi |i - j|}{L}\right)\, .
\label{eq:model}
\end{equation} 
where $S_i \in\{\pm 1\}$ are, for example, Ising spins and the sum ranges 
over all spins in the system. 
In equation (\ref{eq:model}) the $\epsilon_{ij}$ are chosen
from a Gaussian distribution of zero mean and standard deviation
unity, and $c(\sigma)$ is a constant which is chosen such that the
model has a mean-field transition temperature $T_c^{\rm MF} = 1$
(see reference \cite{katzgraber:03} for details). To ensure periodic
boundary conditions the spins are placed on a circular chain of
circumference $L$ and the distance $r_{ij}$ between two spins $i$
and $j$ is thus given by the geometric distance on the circle topology.
The model has a very rich phase diagram in the $d$--$\sigma$ plane,
see figure \ref{fig:dsigma}. Note that here we study the model in
one space dimension, i.e., $d = 1$, which corresponds to the thick
horizontal (white) line in the figure. By changing the power-law
exponent $\sigma$ the universality class as well as the range of the
interactions of the model can be changed continuously for a large range
of system sizes. This has the advantage that the model can be used to
test the applicability of several theoretical predictions made for
the mean-field SK model for finite-range systems.  Furthermore, the
scaling of different algorithms strongly depends on the interaction
range between the spins. While the system is always fully connected,
the range of the interactions and henceforth the effective space
dimension of the model can be tuned as well. Therefore the model is
an ideal benchmark for different optimization algorithms.

\section{Application to spin-glass problems: past, present, and future}
\label{sec:sg}

In what follows an overview over different problems in the field
of spin glasses studied with the one-dimensional Ising chain are
discussed, as well as current and future applications.

\subsection*{Nature of the spin-glass state}

Traditionally, two main pictures have been used to describe the
nature of the spin glass state: replica symmetry breaking (RSB)
\cite{parisi:79,parisi:80,parisi:83,mezard:87} and the droplet picture
\cite{mcmillan:85,fisher:86,fisher:87,fisher:88,bray:86}.  Replica
symmetry breaking predicts that droplet excitations involving a finite
fraction of the spins cost only a finite energy in the thermodynamic
limit. This can be tested by studying the distribution of the spin
overlap $P(q)$ at $q = 0$ \cite{marinari:00,katzgraber:01}. Scaling
relations predict that $P(q = 0) \sim L^{-\theta'}$ with $\theta'
= 0$. Furthermore, the fractal dimension of the excitations is
the same as the space dimension, i.e., $d_s = d$. In contrast,
for the droplet picture one expects $\theta' \neq 0$ and $d - d_s
< 0$, i.e., excitation energies diverge as $E \sim L^{\theta}$
in the thermodynamic limit and the surface of the excitations
is fractal \cite{fisher:86,fisher:87,fisher:88,bray:86}.
Simulations of the one-dimensional Ising spin glass
with power-law interactions have shown that, for system
sizes $L$ considerably larger than in higher-dimensional
models \cite{katzgraber:01}, an intermediate scenario emerges
\cite{katzgraber:03,katzgraber:03f,katzgraber:05d}---known as TNT for
``trivial--nontrivial''---where excitations cost a finite energy but
their surfaces are fractal \cite{krzakala:00,palassini:00} in the
thermodynamic limit.

\subsection*{Ground-state energy distributions in spin glasses}

There has been considerable work in understanding the behavior
of ground-state energy distributions for the  mean-field SK model
\cite{palassini:03a,andreanov:04,boettcher:05,boettcher:05a}. In
particular, it has been shown that the ground-state energy
distributions can possibly be fitted to modified Gumbel distributions
\cite{bramwell:01,koerner:06}. Work on short-range systems---only
possible for small system sizes \cite{bouchaud:03}---suggest Gaussian
ground-state energy distributions in the thermodynamic limit. Thus
the one-dimensional Ising chain offers itself as an ideal model to
test the shape of the distributions when leaving the infinite-range
universality class.

Results \cite{katzgraber:04c} have shown that for $\sigma \le 0.5$,
where the model exhibits infinite-range behavior, the skewness of
the distributions tends to a constant in the thermodynamic limit,
indicating that the data cannot be fitted properly with a Gaussian.
For $\sigma > 0.5$, the skewness decays with a power law of the
system size, indicating that outside the infinite-range region the
ground-state energy distributions become Gaussian in the thermodynamic
limit \cite{wehr:90}. This shows that the infinite-range SK model
shows a singular behavior in this respect \cite{bertin:06}.

\subsection*{Existence of an Almeida-Thouless line in short-range spin
glasses}

There has been an ongoing debate as to
whether short-range spin glasses order in a field or not
\cite{bhatt:85,ciria:93b,kawashima:96,billoire:03b,marinari:98d,houdayer:99,krzakala:01,takayama:04,young:04,jonsson:05a}.
Simulations of three-dimensional Ising spin glasses \cite{young:04}
suggest that the de Almeida-Thouless line \cite{almeida:78}, which
exists for the mean-field SK model, does not exists for realistic
short-range Ising spin glasses. While the aforementioned results found
by studying the two-point correlation length \cite{ballesteros:00}
provide strong evidence that short-range spin glasses do not order
in a field, there are some open questions. First, the system sizes
simulated in reference \cite{young:04} are not very large. Furthermore, it
is unclear if short-range systems above the upper critical dimension
order in a field or not because simulations of high-dimensional spin
glasses are extremely difficult to perform, especially in an externally
applied field.

\begin{figure}[h]
\begin{minipage}{18pc}
\includegraphics[width=20pc]{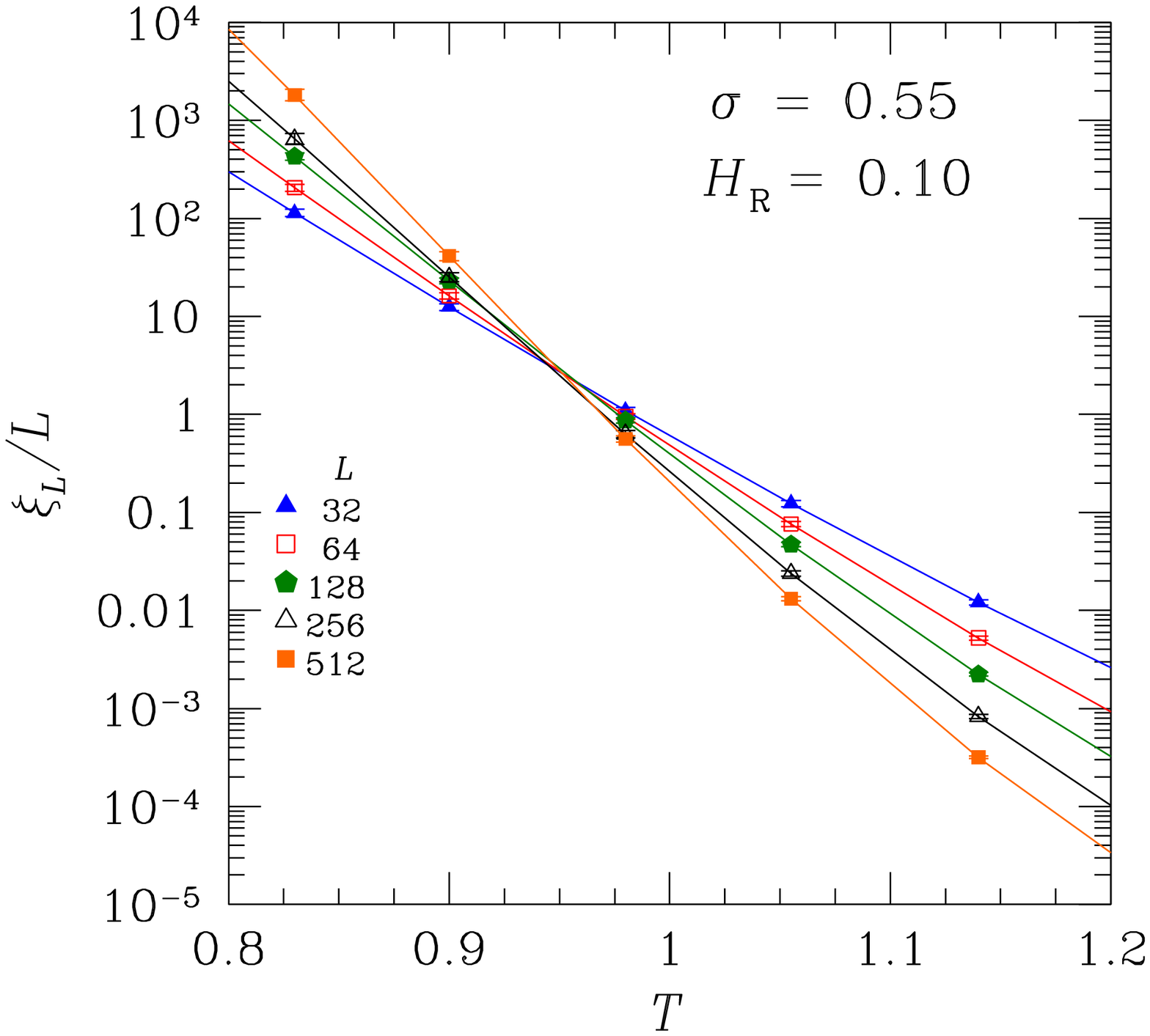}
\vspace*{-1.5cm}
\caption{\label{fig:at1}
Two-point correlation length at finite field for $\sigma =
0.55$. The data cross at $T_{\rm c} \approx 0.95$ suggesting
that there is a spin-glass state in a field. The system is in the
mean-field yet not in the SK universality class. Figure adapted from
reference \cite{katzgraber:05c}.}
\end{minipage}
\hspace{2pc}
\begin{minipage}{18pc}
\includegraphics[width=20pc]{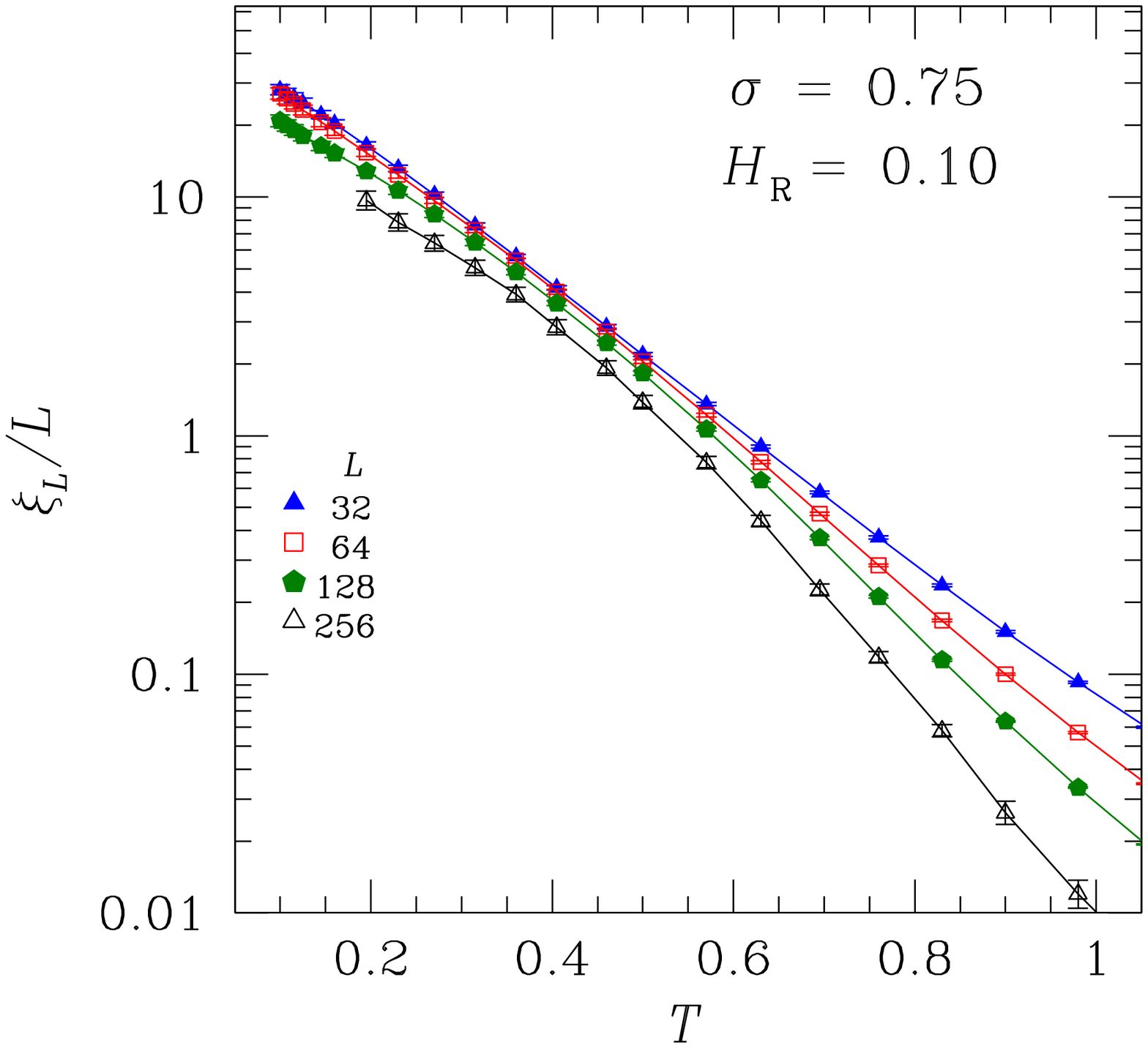}
\vspace*{-1.5cm}
\caption{\label{fig:at2}
Two-point correlation length at finite fields for $\sigma = 0.75$. The
data do not cross even for extremely low $T$ suggesting that there is
no spin-glass state in a field. The system is in the non-mean-field
universality class. Figure adapted from reference \cite{katzgraber:05c}.}
\end{minipage}
\end{figure}

Katzgraber and Young have simulated the one-dimensional Ising
chain in a field for different values of the exponent $\sigma$
\cite{katzgraber:05c} and find that there is no de Almeida-Thouless
line for the range of the power-law exponent corresponding to
a non-mean-field transition in zero field ($\sigma > 2/3$). This
suggests that there is no de Almeida-Thouless line for short-range spin
glasses below the upper critical dimension. In Figs.~\ref{fig:at1}
and \ref{fig:at2} data for a small external field $H_{\rm R} = 0.10$
for $\sigma = 0.55$ (mean-field regime) and $0.75$ (below the upper
critical dimension), are shown. While the data for the correlation
length---which scales as $\xi_L/L = \widetilde{X}\left(L^{1/\nu}[T -
T_{\rm c}(H_{\rm R})]\right)$---cross for $\sigma = 0.55$ suggesting
that there is a spin-glass state at finite fields, this is not the case
for $\sigma = 0.75$ where simulations down to very low temperatures
[$T \ll T_{\rm c}(H_{\rm R} = 0) \approx 0.69(1)$] have been performed.
Data for $\sigma \approx 2/3$ where the one-dimensional Ising chain
changes from the mean-field to the non-mean-field universality class
show marginal behavior (not shown).  Details about
the simulation can be found in reference \cite{katzgraber:05c}.

\subsection*{Field chaos in spin glasses}

The chaotic response of spin glasses to small perturbations in the
temperature, disorder, or externally applied field have been predicted
a  long time ago \cite{mckay:82,parisi:84} and analyzed on the basis
of scaling arguments \cite{fisher:86,bray:87}. Recently, Katzgraber
and Krz\c{a}ka{\l}a have shown that temperature and disorder chaos
in three-dimensional spin glasses can be observed using scaling
laws \cite{katzgraber:07} at low enough temperatures and that both
perturbations seem to share the same scaling functions (although there
was general consensus that disorder chaos is observable in spin glasses
\cite{kondor:89,neynifle:97,neynifle:98,billoire:00,billoire:02}).

We have studied the effects of small perturbations in
the field \cite{ritort:94,billoire:03,sasaki:05} on the
equilibrium state of the one-dimensional Ising spin chain
at low but nonzero temperature. Following previous studies
\cite{ritort:94,neynifle:97,neynifle:98,katzgraber:07,billoire:03} we
study field chaos when the field between two replicas of the system
with the same disorder is shifted by an amount $\Delta H$. To study
the effects of the perturbation we compute the chaoticity parameter
$Q$ given by
\begin{equation}
Q_{\Delta H} = \left[
  \frac{ \langle q^2_{0,\Delta H}\rangle } {\sqrt{ \langle q^2_{0,0}\rangle
      \langle q^2_{\Delta H, \Delta H}\rangle }}\right]_{\rm av} 
\;\; \sim \;\;\;\;
\tilde{Q}[\Delta H/L^{\theta/d - 1/2}] .
\label{eq:Q}
\end{equation}
In equation (\ref{eq:Q}) $q_{a,b} = L^{-1}\sum_iS_i^a S_i^b$ is the
spin overlap between configurations $a$ and $b$ at different
fields, $\langle \cdots \rangle$ represents a thermal average, and
$[\cdots]_{\rm av}$ a configurational average.  Our results show that
the data for the chaoticity parameter $Q$ can be scaled according to
the scaling behavior presented in equation (\ref{eq:Q}) with $d = 1$ and
$\theta \approx 0$ for {\em very low temperatures} $T = 0.1 \ll T_c$.
These {\em preliminary} results for small system sizes $L$ and few
values of $\Delta H$ suggest that field chaos could be present in
short-range as well as long-range spin glasses.

\begin{figure}[h]
\includegraphics[width=20pc]{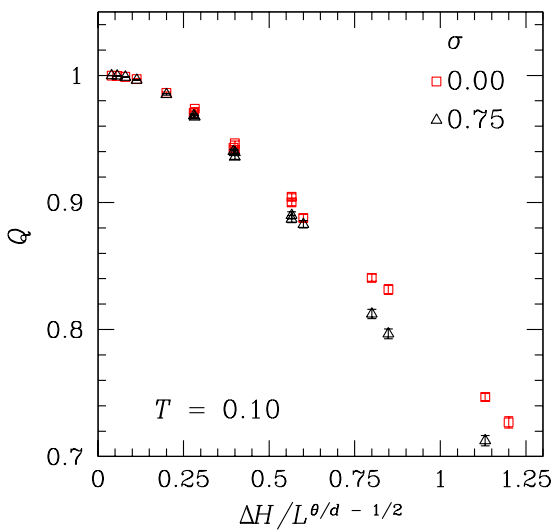}\hspace{2pc}
\begin{minipage}[b]{16pc}\caption{\label{fig:scale_Q}
Chaoticity parameter $Q$ as a function of $\Delta H/L^{\theta/d -
1/2}$ for $\sigma = 0.00$ and $0.75$ and (random) fields of strength
$\Delta H = 0.01$, $0.05$, $0.07$, $0.10$, and $0.15$ for $T = 0.10$
($L = 16$, $32$, $64$, and $128$).  The data for different fields
collapse onto universal curves for the different values of $\sigma$
and $\theta \approx 0$. Note that for the SK model the data should
collapse with $\Delta H /L^{3/8}$ \cite{billoire:03}. This is not the case for the
present results (work in progress \cite{katzgraber:07b}).
\vspace*{1cm}
}
\end{minipage}
\end{figure}

\vspace*{-1.5cm}

\subsection*{Ultrametricity in spin glasses}

One of the cornerstones of the Parisi solution of the mean-field
Sherrington-Kirkpatrick model is the concept of ultrametricity
\cite{mezard:84}, but it is unclear if realistic short-range
spin glasses exhibit this property in the low-temperature phase
\cite{contucci:07,joerg:07}.  Ultrametricity can be described as
follows: Consider an equilibrium ensemble of states at $T < T_{\rm
c}$ and pick three, $\rho$, $\mu$ and $\nu$, at random. These indices
are associated with the states $S^\rho$, $S^\mu$ and $S^\nu$. Order
them so that the overlap $q_{\mu\nu} = L^{-1}\sum S_i^\mu S_i^\nu$
between them satisfies $q_{\mu\nu}\geq q_{\nu\rho}\geq q_{\mu\rho}$.
Ultrametricity means that in the thermodynamic limit, we obtain
$q_{\nu\rho}=q_{\mu\rho}$ with probability $1$. Recent results on small
three-dimensional systems suggest that short-range spin glasses do not
possess this characteristic of the mean-field model \cite{hed:03},
although opposing opinions \cite{franz:00} exist. Capitalizing on
the success of the one-dimensional Ising spin glass with power-law
interactions in elucidating different properties of spin glasses we
have studied ultrametricity in spin glasses for different exponents
$\sigma$.

\begin{figure}[h]
\begin{minipage}{18pc}
\includegraphics[width=14pc]{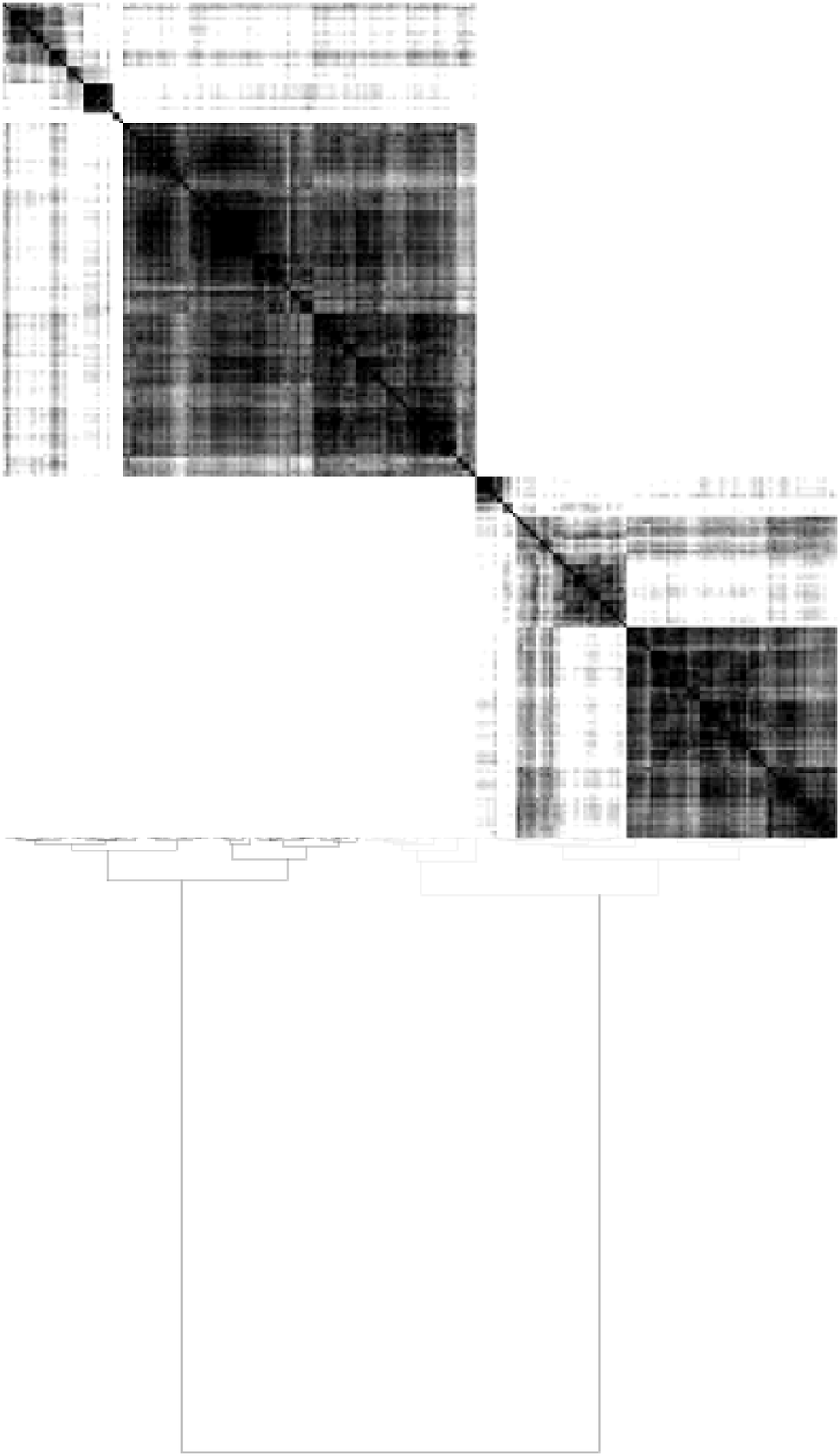}
\caption{\label{fig:ul0.00}
Dendrogram for the one-dimensional Ising chain for $\sigma = 0.00$
(SK model).  Darker color correspond to closer distances. Data for $L =
512$ and $T = 0.20$. For details see the main text.}
\end{minipage}
\hspace{2pc}
\begin{minipage}{18pc}
\includegraphics[width=14pc]{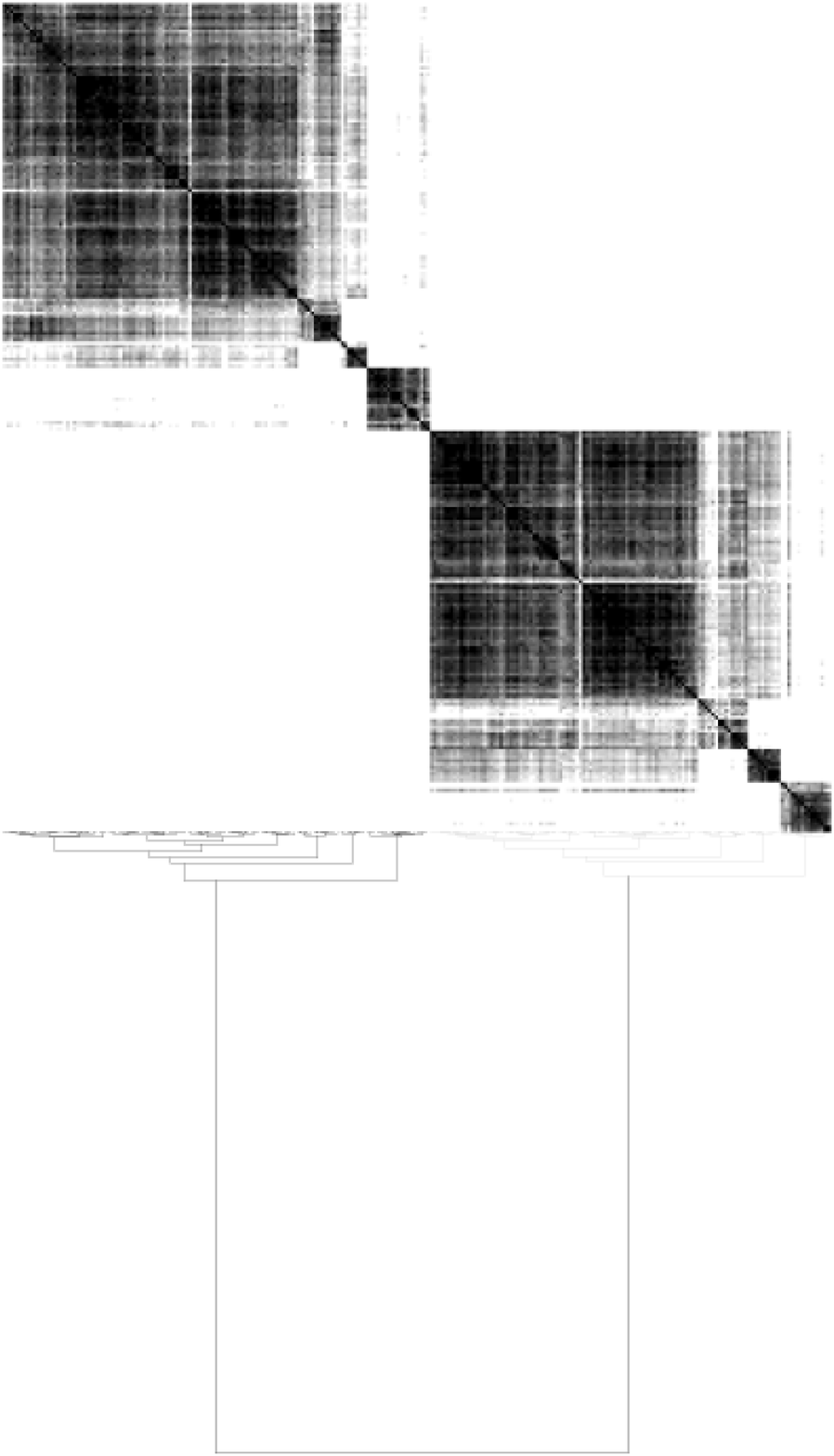}
\caption{\label{fig:ul0.75}
Dendrogram for the one-dimensional Ising chain for $\sigma = 0.75$
(long-range universality class). Darker color correspond to closer
distances. Data for $L = 512$ and $T = 0.20$.}
\end{minipage}
\end{figure}

In Figs.~\ref{fig:ul0.00} and \ref{fig:ul0.75} we show preliminary
sorted dendrograms using Ward's method \cite{ward:63} for the SK model
and the one-dimensional Ising chain for $\sigma = 0.75$ (data for $L =
512$, $T = 0.20$). Displayed are equilibrium states in configuration
space. The distance between the states in the distance matrix is
color coded (darker color corresponds to closer distances). The
states are sorted linearly, where the order is determined from the
hierarchical clustering \cite{ward:63} of the tree-structure of the
underlying dendrogram (bottom panel in the figures). The clustering
procedure starts with $L$ clusters which contain one state and
the two closest lying clusters are merged (joining lines in the
dendrogram). This procedure is repeated until one large cluster is
obtained.  There is clearly structure in the dendrograms and a valid
hierarchical clustering corresponds to an ultrametric structure of
space \cite{rammal:86}.  Note that this is not the case for $T > T_c$
where the dendrograms show no structure at all (not shown). To further
strengthen the aforementioned results, a finite-size scaling analysis
of the data will be performed (work in progress \cite{katzgraber:07a}).

\subsection*{Local-field distributions in spin glasses}

The distribution of local fields $P(h = \sum_j J_{ij}S_j,T)$ at a
temperature $T$ has been of interest since the early days of spin
glasses \cite{marshall:60,klein:63,thouless:77}. In particular,
the behavior of the mean-field SK model is well understood
\cite{thomsen:86,oppermann:05,pankov:06,oppermann:07}.  On the other
hand, there has been little work for short-range Edwards-Anderson
spin glasses since these can only be studied numerically.

It has been shown for the mean-field SK model that $P(h) \sim a
|h|$ for $h \rightarrow 0$ and $T = 0$ in the thermodynamic limit.
This suggests that spins with zero local field exist, i.e., domain
walls can move freely at no energy cost in the system.  Simulations for
short-range finite-dimensional systems and intermediate system sizes
have shown (unpublished work \cite{boettcher:07}) that $P(h) \sim c +
a |h|$ for space dimensions $d \ge 2$ with possibly a finite value of
$c$ in the thermodynamic limit. We have calculated the local field
distribution for the one-dimensional Ising spin chain for different
values of $\sigma$. Extrapolating the data to $T = 0$ we can study
the behavior of $P(h = 0, T = 0) = c$ as a function of the system size
$L$. Our results show that while for the SK model $c = 0.006(9)$, i.e.,
$P(h = 0, T = 0)|_{L \rightarrow \infty} \rightarrow 0$, for $0.5 <
\sigma < \infty$ finite values of $c$ in the thermodynamic limit
are obtained, e.g., for $\sigma = 0.75$ $c = 0.021(1)$. This again
highlights the singular behavior of the SK model \cite{boettcher:07}.

\subsection*{Future directions}

So far the model has primarily been used to study properties
of Ising spin glasses. A possible future direction would be to
study versions of the model with different spin symmetries or
dilution (which might allow the simulation of larger systems).
For example, there has been considerable interest in the nature
of the spin-glass state of the three-dimensional Heisenberg
spin glass \cite{kawamura:92,lee:03,hukushima:05,campos:06}. In
particular, it is unclear if spin and chirality degrees of freedom
decouple. Recently, a one-dimensional Heisenberg chain with power-law
interactions \cite{matsuda:07} has been studied in an attempt to
answer this problem. There, simulations for $\sigma = 1.1$ where
$T_c$ for the spin-glass sector is zero have been interpreted as a
spin-chirality decoupling scenario since the chiralities showed a
nonzero transition temperature. The model could also be extended to
study XY spins.  Furthermore, nonequilibrium properties in spin glasses
\cite{montemurro:03} can also be studied for large system sizes.
Finally, modifications of the Hamiltonian might be used to address
problems in different fields, e.g., a $p$-spin version \cite{moore:06}
of the one-dimensional Ising spin chain to study structural glasses
(work in progress).

\section{Algorithm benchmarking}
\label{sec:alg}

The development (and testing) of algorithms to study systems with
complex energy landscapes \cite{hartmann:01,hartmann:04} plays a
crucial role in the field of statistical mechanics of disordered
systems, as well as many interdisciplinary applications to other
fields. We discuss some examples below.

\begin{figure}[h]
\begin{minipage}{18pc}
\includegraphics[width=20pc]{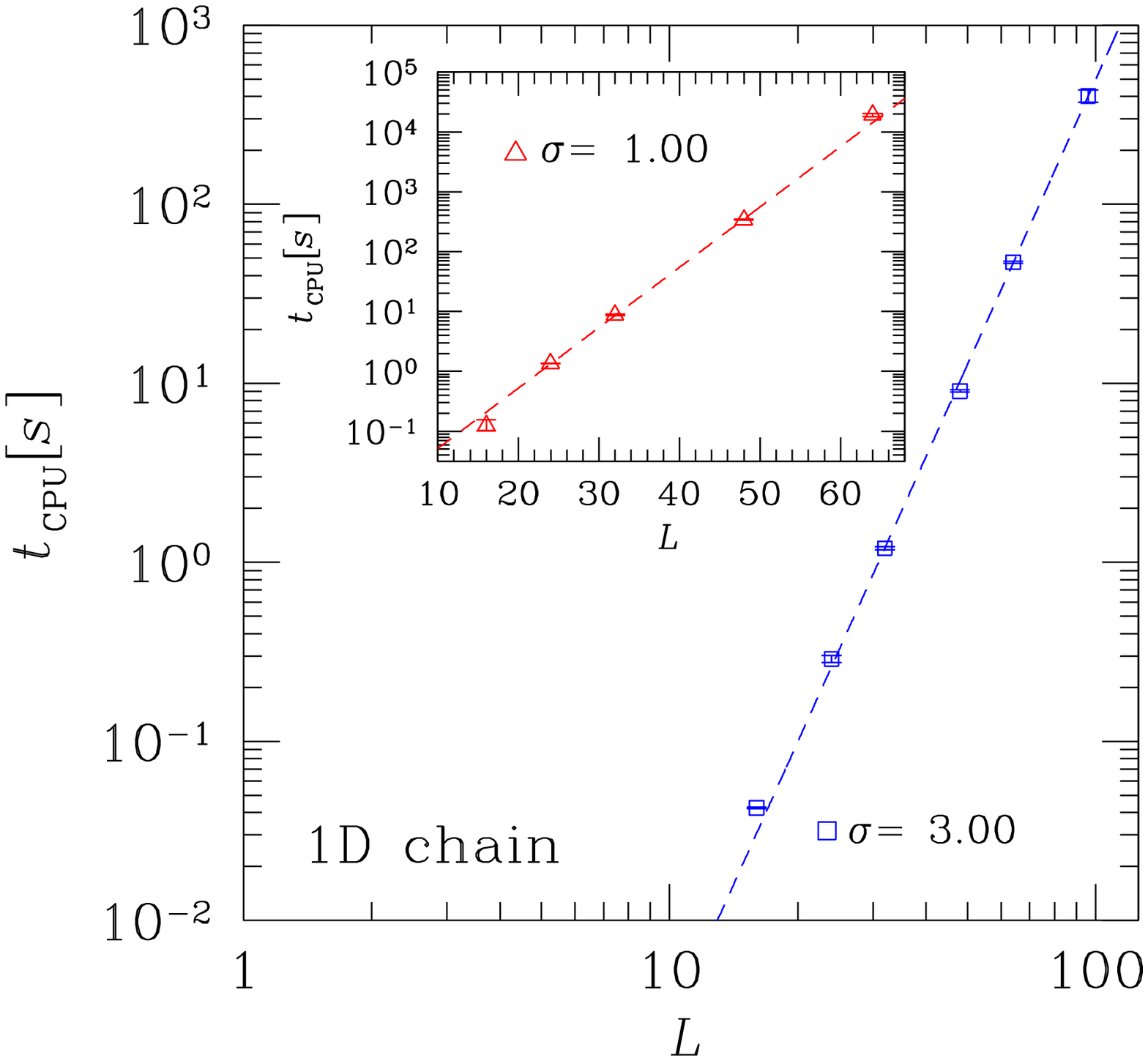}
\vspace*{-1.5cm}
\caption{\label{fig:cpu_L}
Mean CPU time in seconds ($t_{\rm CPU}$) for determining a ground state
of the one-dimensional Ising chain as a function of the chain length
$L$ for different exponents $\sigma$ using the branch, cut \& price
algorithm. For $\sigma = 3.0$ (main panel) the CPU time increases $\sim
L^{5.3}$, whereas for $\sigma \lesssim 2.0$ the CPU time increases
exponentially (see inset for $\sigma = 1.0$).  The dashed lines are
guides to the eye.  Figure adapted from reference \cite{katzgraber:05d}.
}
\end{minipage}
\hspace{2pc}
\begin{minipage}{18pc}
\includegraphics[width=20pc]{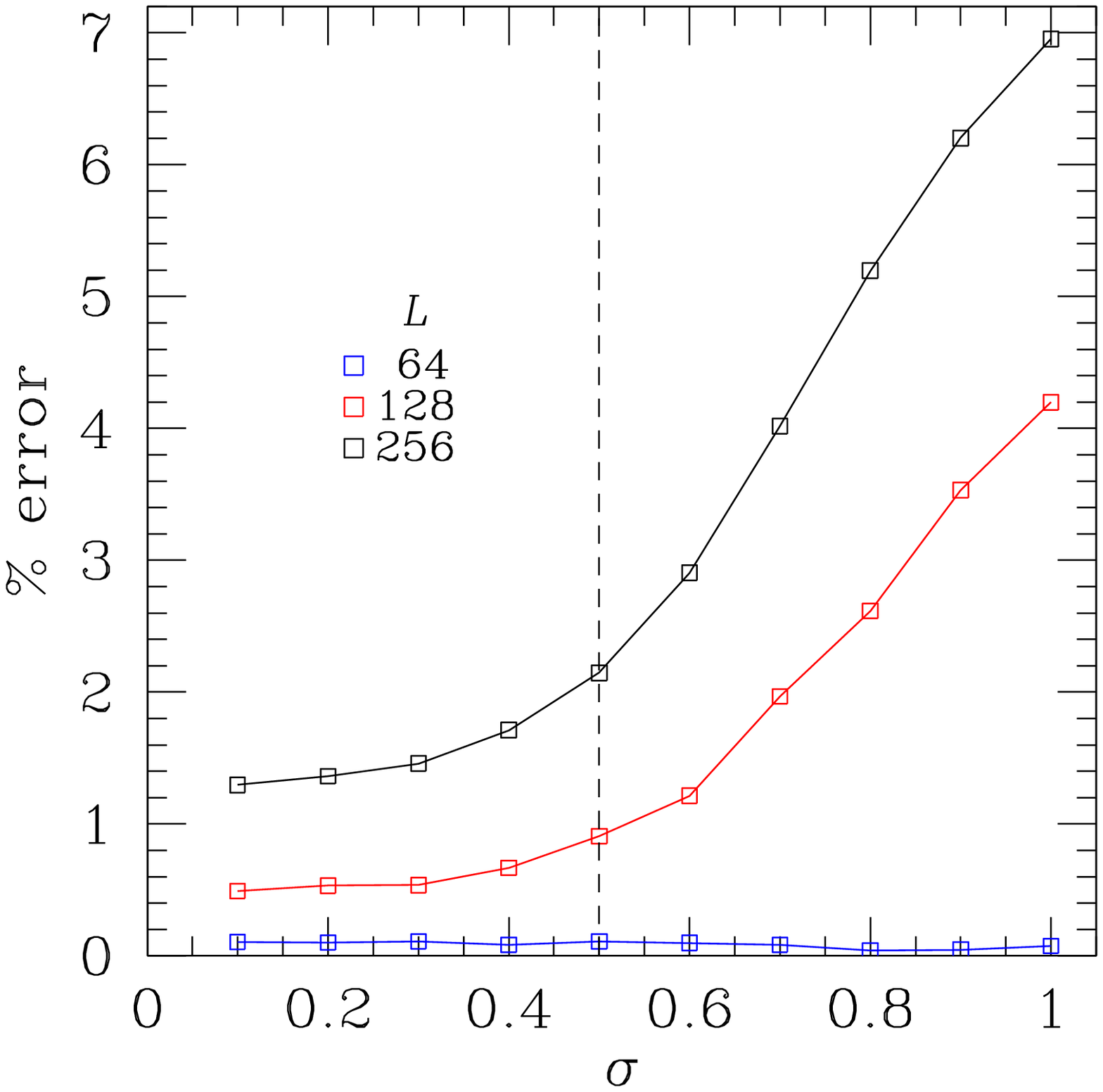}
\vspace*{-1.5cm}
\caption{\label{fig:eo}
Percentage error in the ground-state energies obtained with hysteretic
optimization with respect to exact ground states obtained with other
approaches as a function of the exponent $\sigma$ for different
system sizes. The algorithm works relatively well for $\sigma \lesssim
0.5$ (vertical dashed line)
whereas for larger values of $\sigma$, where the model is not
infinite ranged the error increases considerably. Figure adapted
from reference \cite{goncalves:07}.}
\end{minipage}
\end{figure}

In the past \cite{katzgraber:05d} we have used exchange Monte Carlo
\cite{hukushima:96} to obtain ground-state energies for spin-glass
systems \cite{moreno:03}.  In this technique, one simulates several
copies of the system at different temperatures, and, in addition
to the usual local Monte Carlo moves, one performs global moves in
which the temperatures of two copies with adjacent temperatures are
exchanged to overcome energy barriers in complex energy landscapes.
By choosing a low enough minimal temperature, the ground state of the
system can be probed.  Interestingly, the algorithm works well for
small values of the exponent $\sigma$, whereas for large $\sigma$
exchange Monte Carlo does not equilibrate in reasonable amounts
of time---possibly because it is difficult to push domain walls
out of the system. Conversely, the branch, cut \&  price algorithm
\cite{barahona:88,liers:04,juenger:95} works best for large $\sigma$
values (see figure \ref{fig:cpu_L}), i.e., in this case complementing
exchange Monte Carlo.

Recently, the hysteretic \cite{pal:06} and extremal
\cite{boettcher:01,boettcher:05} optimization methods have been
introduced to heuristically estimate ground-state energies of spin
glasses. Hysteretic optimization successively demagnetizes the
system at zero temperature with some additional shake-ups until
states close to the ground state are reached.  In a recent project,
Gon{\c c}alves and B\"ottcher \cite{goncalves:07} have studied the
efficiency of hysteretic optimization when computing ground-state
energies of the one-dimensional Ising chain as a function of the
exponent $\sigma$. Their results clearly show that the method works
best for infinite-range models ($\sigma \le 0.5$) where avalanches
in the hysteresis loops proliferate easily. Once the system is not
infinite ranged, avalanche sizes are small and the algorithm is trapped
(not shown).  This shows that while hysteretic optimization is a fast
method for fully-connected models such as the SK model or the traveling
salesman problem, it is not efficient for short-range spin glasses.

\section{Conclusions}
\label{conclusions}

By using a one-dimensional spin glass with power-law interactions we
have been able to study a variety of open questions in the field of
spin glasses. The model has two main advantages over conventional
higher-dimensional models: larger system sizes can be studied and
the universality class of the model can be tuned by changing the
power-law exponent of the interactions. The different results obtained
show that there is urgent need for a better theoretical description
of short-range spin glasses. While the droplet model and replica
symmetry breaking describe certain properties of spin glasses well,
neither of both theories is able to deliver a full account of all
properties of short-range systems.

Furthermore, the model serves as a strong benchmark for different
optimization algorithms: Because the range of the interactions can
be tuned, the applicability of algorithms to different models in
different universality classes can be tested.

\ack{I would like to thank W.~Barthel, S.~B\"ottcher, B.~Gon{\c
c}alves, A.~K.~Hartmann, M.~J\"unger, M.~K\"orner, F.~Krz\c{a}ka{\l}a,
F.~Liers, D.~Sherrington and A.~P.~Young for fruitful collaborations
and to I.~A.~Campbell and T.~J\"org for discussions. In particular,
I would like to thank B.~Gon{\c c}alves and S.~B\"ottcher for sharing
their data presented in figure \ref{fig:eo} prior to publication.
Part of the simulations have been performed on the Asgard, Gonzales
and Hreidar clusters at ETH Z\"urich.  This work has been supported
by the Swiss National Science Foundation under Grant No.~PP002-114713.}

\section*{References}

\bibliography{refs,comments}

\end{document}